\documentclass[aps,prb,preprint]{revtex4}
\usepackage{graphicx,psfrag}
\usepackage{graphicx} % Include figure files
\usepackage{bm}% bold math
\usepackage{booktabs}% tabs
\usepackage{amsmath,amsthm,amssymb}
\newlength{\minitwocolumn}
\usepackage{color}% color
\begin{document}

\title{Ultrafast magnetic vortex core switching driven by topological inverse Faraday effect}
\author{Katsuhisa Taguchi,$^1$ Jun-ichiro Ohe,$^2$ and  Gen Tatara$^1$ }  
\affiliation{$^1$Department of Physics, Tokyo Metropolitan University, Hachioji, Tokyo 192-0397 Japan \\
$^2$Department of Physics, Toho University, Funabashi, Chiba 274-8510 Japan
}
\date{\today}
\begin {abstract} 
We present a theoretical discovery of an unconventional mechanism of inverse Faraday effect (IFE) which acts selectively on topological magnetic structures. The effect, topological inverse Faraday effect (TIFE), is induced by spin Berry's phase of the magnetic structure when a circularly polarized light is applied.
Thus a spin-orbit interaction is not necessary unlike in the conventional IFE. 
We demonstrate by numerical simulation that TIFE realizes ultrafast switching of a magnetic vortex within a switching time of 150 ps without magnetic field.
\end{abstract}

\maketitle

Ultrafast magnetization switching without magnetic field is required from the view point of spintronics application for achieving high speed read-write processes in non-volatile memories.
Current-induced magnetization dynamics is one of the possible phenomena for realizing fast processes.
By applying a charge current to the magnetic domain wall, the domain wall moves with the velocity of $\sim 100$ m/s \ \cite{rf:Parkin08}, which corresponds to a switching time of $1$ ns when the device size is $100$ nm.
A magnetic vortex core is another candidate of a memory device where the polarization direction of the core is used as information.
If one uses a magnetic field, switching of a vortex core can be carried out within 140 ps \ {\cite{rf:Xiao06}}, but use of magnetic field has a disadvantage in realizing high density memories.
A switching by field-driven spin wave excitation was recently carried out and switching time of 200 ps was obtained \cite{rf:Hermann11}.  
In a switching of a vortex core carried out by applying an electric current pulse,
the switching time was $\sim 20$ ns \   \cite{rf:Yamada06}.
There are, however, two difficulties to apply the current-induced vortex core switching to memory devices.
First, a high current density ($\sim$10$^{12}$ A/m$^2$) is needed.
Heating effect due to high current density causes an unstable operation.
Second, a slow processional motion of the core is needed before the core switching occurs.
This transient motion makes it difficult to predict when the core switching occurs.
It is therefore desired to find other mechanisms for realizing fast vortex core switching.

Recent studies using femtosecond laser pulses revealed fast magnetic switchings\cite{rf:Beaurepaire96,rf:Ogasawara09,rf:Radu11}.
A subpicosecond demagnetization of Ni was induced by a heating effect of a femtosecond laser pulse\cite{rf:Beaurepaire96}. 
In Ref. 6, a magnetic field was applied besides a laser pulse to realize magnetization reversal in TbFeCo.
Recently, thermal reversal of a ferrimagnet GdFeCo by applying a laser pulse was carried out, and reversal mechanism was explored\cite{rf:Radu11}.
Use of circularly polarized light \cite{rf:Stanciu07,rf:Vahaplar} has a great advantage in realizing magnetization switching in terms of all-optical methods without magnetic field.
Circularly polarized light induces a magnetization switching by inverse Faraday effect (IFE) \cite{rf:Pitaevskii61,rf:Pershan66}.
An effective magnetic field generated by the polarized light via conventional IFE turned out to reach 20 T for a strong laser pulse, and such effective field makes possible a switching in a time scale of  60 ps \  \cite{rf:Vahaplar}.
The conventional IFE emerges in the presence of uniform magnetization and spin-orbit interaction.
To decrease the switching time, therefore, one needs to use heavy elements to enhance the spin-orbit interaction. 

Recent studies on the anomalous Hall effect have pointed out the similarity of spin-orbit interaction and topological magnetic structures \cite{rf:Ye99,rf:Nagaosa10}.
In fact, topological structures carrying spin Berry's phase couple the coordinate space with spin space just like spin-orbit interaction does.
One can thus expect that the role of the spin-orbit interaction in IFE  can be replaced by spin Berry's phase of topological magnetic structures.
A great advantage of using the topological magnetic structure is that the effective flux of the Berry's phase can be as large as one flux quanta ($\frac{h}{e}=4\times 10^{-15}$ Wb) per unit area of lattice.
For a vortex with core radius of 10 nm, the effective field of spin Berry's phase thus reaches 6 T.

In this letter, we show that there is indeed an unconventional IFE acting selectively on topological magnetic structures 
with spin Berry's phase or chirality, such as magnetic vortices and skyrmions\cite{rf:skyrmion}.
By analytical calculation based on the nonequilibrium Green's functions, we derive a general formula of the effective magnetic field induced by the incident polarized light and the Berry's phase of magnetization structure.
We call the present Berry's phase induced IFE as topological IFE (TIFE).
The effect is useful to control individual vortex or skyrmion without touching the background, and has advantage for designing devices based on topological magnetic structures.
Furthermore, we demonstrate that the ultrafast magnetic vortex core switching in 150 ps can be achieved in magnetic disk by using a numerical simulation.

The problem we are going to solve is to construct a vector representing the effective magnetic field induced by TIFE,  $\bm{H}_{{\rm{TIFE}}}$, from three vectors, $\bm{L}$ proportional to the helicity of incident light, direction of local spin $\bm{n}$ and Berry's phase field, $\bm{\Phi}$.
From a symmetry argument, we have three possibilities; $\bm{H}_{{\rm{TIFE}}} \propto \bm{n} \left( \bm{\Phi} \cdot \bm{L}\right)$, $\bm{L} \left( \bm{\Phi} \cdot \bm{n}\right)$, and $\bm{\Phi} \left( \bm{L} \cdot \bm{n}\right)$.
Which coupling emerges in reality is answered by a microscopic calculation. 
We will show that $\bm{H}_{{\rm{TIFE}}} \propto \bm{n} \left( \bm{\Phi} \cdot \bm{L}\right)$ is the correct answer. 
The effective field indicates that the energy density of the system when a circularly polarized light is injected is 
$E_{{\rm{TIFE}}} =- g \mu_{\rm{B}} S  \bm{H}_{{\rm{TIFE}}} \cdot \bm{n} \propto \left( \bm{\Phi} \cdot \bm{L}\right)$, where $g$ is the Land$\acute{{\rm{e}}}$ factor, $\mu_{\rm{B}}$ is the Bohr magneton, and  $S$ is the magnitude of local spin.  
The angular momentum of light is therefore coupled to the effective magnetic flux of the spin Berry's phase, in the same manner as the orbital ($\bm{\ell}$) or spin ($\bm{S}$) angular momentum couples to a magnetic field ($\bm{H}$) as $\bm{\ell} \cdot \bm{H} $ and $\bm{S} \cdot \bm{H} $.

Let us derive an analytical expression for TIFE by a microscopic calculation. 
We consider conduction electrons interacting with a local spin structure by a $s$-$d$ type interaction. The electric field of the incident light is included to the second order to describe the circular polarization.
We consider a light in the THz regime. This simplifies theoretical study greatly, since then the electron excitation is limited only very close to the Fermi level. Nevertheless, the results obtained in the present paper are expected to be qualitatively extended to the case of the visible light, as suggested by qualitatively the same results for visible light and THz light in the case of conventional IFE\cite{rf:KT11}.  

%The spin structure is represented by static classical vectors, $\bm{S}(\bm{r}) = S\bm{n}(\bm{r})$, where $\bm{n}$ is the unit vector representing the direction and $S$ is the magnitude. 

The spin structure is represented by static classical vectors, $\bm{S}(\bm{r}) = S\bm{n}(\bm{r})$, where $\bm{n}$ is the unit vector representing the direction.
The $s$-$d$ interaction between the conduction electron and localized spins reads
\begin{align} 
\mathcal{H}_{\rm{sd}}= - \Delta \int d^3\bm{r} \bm{n}\cdot \left(c^\dagger \bm{\sigma} c \right),
\end{align} 
where $c\equiv \left( c_{\uparrow}, c_{\downarrow}\right)$ and $c^\dagger$ are the annihilation and creation operators of the conduction electrons, respectively (indices $\uparrow$ and $\downarrow$ represent spin).  
We consider a strong coupling case, where the $s$-$d$ exchange splitting, $\Delta$, is large. The conduction electron spin thus follows almost perfectly the localized spin structure, namely it is in the adiabatic regime\cite{rf:Tatara08}.
The electric field of applied light is represented by 
$\bm{E}={\rm{Re}}\left[ \bm{\mathcal{E}} e^{i(\bm{Q}\cdot\bm{r}-\Omega t)}\right]$, where  $\mathcal{E}$ is  complex amplitude, and $\bm{Q}$ and $\Omega$ are wave vector and frequency, respectively.

We use a spin gauge transformation to extract low energy excitations. \cite{rf:Tatara08}
Defining an electron operator $a$ in the gauge transformed frame as $a(\bm{r}) = U(\bm{r}) c(\bm{r})$, where $U$ is a 2 $\times$ 2 unitary matrix, we diagonalize the $s$-$d$ interaction as $\mathcal{H}_{\rm{sd}}= -\Delta \int d^3\bm{r} a^\dagger \sigma_z a$.
This diagonalization is accomplished by choosing $U=\bm{m}\cdot \bm{\sigma}$ with $\bm{m}= \left( \sin{\frac{\theta}{2}}\cos\phi, \sin{\frac{\theta}{2}}\sin\phi, \cos{\frac{\theta}{2}} \right)$, where $\theta$ and $\phi$ are the polar coordinates of $\bm{n}$ \ \cite{rf:Tatara08}.
By the above local gauge transformation, an SU(2) gauge field, $A^\alpha_{{\rm{S}},i} \equiv -\frac{i}{2}{\rm{Tr}}\left[\sigma^\alpha U^\dagger \partial_i U \right]$ ($\alpha=x, y, z$), arises which couples to the spin current, defined as 
$j_{{\rm{S}},i}^\alpha \equiv \frac{\hbar^2}{2mi}[ (\partial_\mu a^\dagger) \sigma^\alpha  a -a^\dagger \sigma^\alpha \partial_\mu a - i A^\alpha_{{\rm{S}},i} a^\dagger a]$.
The total Lagrangian for the conduction electron thus reads 
\begin{align}\notag
\mathcal{L} = & \int d^3 \bm{r} a^\dagger 
	\left( 
	i\hbar \frac{\partial }{\partial t} + \frac{\hbar^2}{2m}\nabla^2 +\epsilon_{\rm{F}} - \Delta \sigma^z 
	\right)  a \\
		   &-  \int d^3 \bm{r} \left[ \sum_{i\alpha} \left( A_{{\rm{S}}, i}^\alpha j_{{\rm{S}}, i}^\alpha + A_i j_i \right)\right], 
\end{align}
where $j_i = \frac{e\hbar}{2mi}[ a^\dagger  \partial_i a - (\partial_i a^\dagger) a  - i\frac{e}{\hbar}A_i a^\dagger a + 2i A^\alpha_{{\rm{S}},i} (a^\dagger \sigma^\alpha a)]$ is a charge current, 
which couples to a vector potential of light, ${\bm A}$, defined as ${\bm E}=-\partial_t {\bm A}$.

The TIFE is studied by calculating the local spin density induced by the incident light, 
$\bm{s}(\bm{r}) \equiv \langle c^\dagger \bm{\sigma} c \rangle$.
In terms of the lesser Green's function\cite{rf:book1} for the gauge-transformed operator,  $G^< (\bm{r},t,\bm{r}',t') \equiv \frac{i}{\hbar} \langle a^\dagger(\bm{r}',t') a(\bm{r},t) \rangle$, $i$-component ($i=x, y, z$) of the spin density reads 
$s_i (\bm{r}) = -i\hbar \sum_j \mathcal{R}_{i j} {\rm{Tr}}\left[ \sigma^j G^< (\bm{r},t,\bm{r},t)\right]
$, where $\mathcal{R}_{ij}\equiv 2m_i m_j -\delta_{ij}$ represents components of a 3 $\times$ 3 rotation matrix. The effective magnetic field induced by TIFE thus
 \begin{equation}
H_{\rm{TIFE},i}= -i\hbar \frac{\Delta}{g\mu_{\rm{B}} S} \sum_j \mathcal{R}_{i j} {\rm{Tr}}\left[ \sigma^j G^< (\bm{r},t,\bm{r},t)\right].
\end{equation}

\begin{figure}\centering
\includegraphics[scale=0.45]{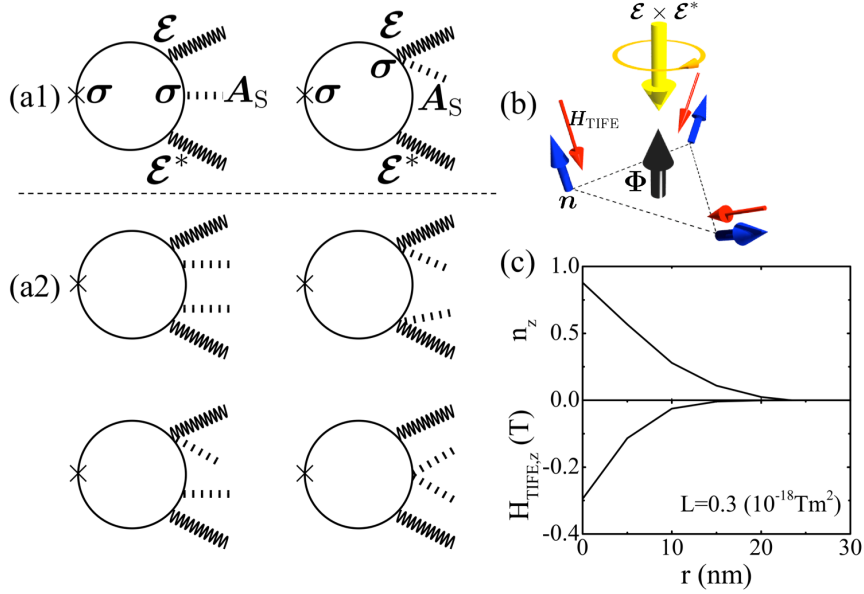}
\caption{(color online) (a) Feynman diagrams representing the effective magnetic field generated by an topological inverse Faraday effect (TIFE). Solid line represents electron's Green's function, and wavy line represents the electric field ($\bm{\mathcal{E}}$) of the incident circularly polarized light. Contributions (a1) and (a2) are linear and second order in spin gauge field ($\bm{A}_{\rm{S}}$, denoted by dotted line), respectively. 
(b) Schematic depiction of the TIFE induced by a circularly polarization, $\bm{\mathcal{E}}\times \bm{\mathcal{E}}^*$, and spin chirality, $\bm{\Phi}$. The induced magnetic field, $\bm{H}_{\rm{TIFE}}$, is parallel or anti-parallel to the local spin direction, $\bm{n}$.
(c) Profile of effective magnetic field, $H_{\rm TIFE}$ (Eq. (\ref{HTIFE})), 
induced for a vortex as a function of distance from the core center, $r$.
The light intensity is $L=0.3\times 10^{-18}$ Tm$^2$. 
}
\end{figure}

The lesser Green's function is evaluated by a standard perturbative method\cite{rf:book1}.
The lowest-order contributions to TIFE are diagrammatically shown in Fig. 1 (a). 
The vector potential of the incident light, ${\bm A}$,  and the spin gauge field, ${\bm A}_{\rm{S}}$, are included to the second order. The contributions are calculated expanding Green's functions with respect to the frequency of THz light, $\Omega$, and to the wave vector of the spin gauge field, $\bm{q}$, and we keep the lowest contributions. 
We note here that the contributions linear in $A_{\rm{S}}$ (the first two diagrams of Fig. 1 (a)) and the one second order in $A_{\rm{S}}$ (the last four diagrams of Fig. 1 (a)) result in the contribution of the same form proportional to  $\epsilon_{j\ell m} \partial_\ell A_{{\rm{S}}, m}^\beta$ owing to the identity satisfied by the spin gauge field\cite{rf:Tatara08}, $\partial_\ell A_{{\rm{S}}, m}^\beta - \partial_m A_{{\rm{S}}, \ell}^\beta - \epsilon_{\beta \gamma \delta}  A_{{\rm{S}}, \ell}^\gamma A_{{\rm{S}}, m}^\delta = 0$.
After a straightforward calculation, we obtain the lowest contribution to the effective field as 
\begin{align}\label{eq:IFE-SC}
H_{{\rm{TIFE}},i} = \sum_{j} \alpha_{ij} \left(\bm{\mathcal{E}} \times \bm{\mathcal{E}}^* \right)_j,
\end{align}
where $\alpha_{ij} = i\tilde{\alpha} \sum_{\ell m} \mathcal{R}_{iz} \epsilon_{j \ell m} \partial_\ell A_{{\rm{S}}, m}^z $,
and $\tilde{\alpha} =  \frac{1}{g\mu_{\rm{B}} S} \frac{ \pi e^2  \hbar \Omega }{m^2 } \left( \nu_{\uparrow}+\nu_{\downarrow} \right) \left( \tau_\uparrow + \tau_\downarrow \right)^2 \left(\epsilon_{\rm{F} \downarrow}\tau_\downarrow^2  - \epsilon_{\rm{F} \uparrow}\tau_\uparrow^2 \right)$.
Here $\nu_{\uparrow(\downarrow)}$ is the density of states, $\tau_{\uparrow(\downarrow)}$ is the electron's lifetime, and $\epsilon_{\rm{F} \uparrow(\downarrow)}$ is the Fermi energy, which depend on the spins (up or down), respectively. 
Coefficient $\alpha_{ij}$ is simplified using spin chirality field defined as 
$\Phi_i = \sum_{\ell m} \epsilon_{i\ell m} \partial_\ell A_{\rm{S},m}^z$ and by noting $\mathcal{R}_{iz}=n_i$ 
as \cite{rf:Tatara08}
\begin{align}\label{eq:a-SC}
\alpha_{ij}  = i \tilde{\alpha} n_i \Phi_j.
\end{align} 
In terms of vector $\bm{n}$, scalar chirality of a spin texture is $\Phi_i= \sum_{jk}\epsilon_{ijk} \bm{n}\cdot\left(\partial_j \bm{n} \times \partial_k \bm{n} \right)$,
where the direction of $\bm{\Phi}$ is perpendicular to the plane the spins lie.
Equation (\ref{eq:a-SC}) indicates that an effective field arises when the magnetic structure has a finite spin chirality ($\bm{\Phi}$) like vortices and skyrimons. 
The effective field is also written as
\begin{equation}
\bm{H}_{\rm{TIFE}}= \left(\bm{L}  \cdot \bm{\Phi} \right)\bm{n} ,\label{HTIFE}
\end{equation}
 where 
\begin{align}
\bm{L} & =  \tilde{\alpha} i \left(\bm{\mathcal{E}} \times \bm{\mathcal{E}}^* \right),
\end{align}
is proportional to the light intensity and helicity.
As seen in Eq. (\ref{HTIFE}), the direction of the field is parallel or anti-parrallel to the local spin direction, $\bm{n}$ (see Fig. 1(b)).
The magnitude of the effective field is determined by the angle between the helicity of the light, $i\left(\bm{\mathcal{E}}\times \bm{\mathcal{E}}^* \right)$, and the spin chirality, $\bm{\Phi}$.
Therefore, the effective field becomes large when a light is emitted along $z$ direction when a vortex structure is formed in the $x$-$y$ plane.

%%%%
\begin{figure}\centering
\includegraphics[scale=0.35]{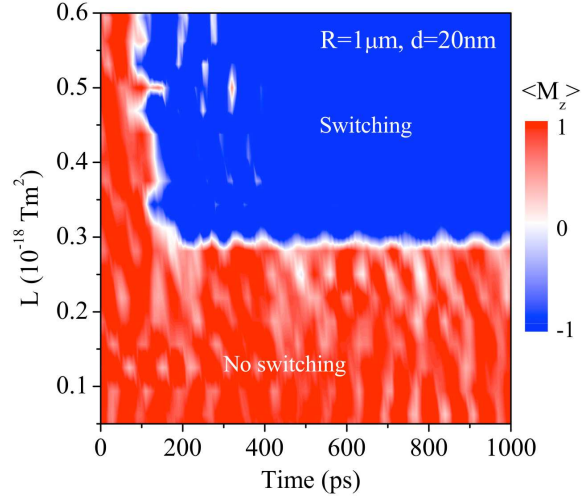}
\caption{ (color online).  
Phase diagram of vortex core reversal for a disk with radius of 1 $\mu $m and thickness of $20$ nm at $T= 100$ K. $L$ is proportional to the laser intensity.
Region colored by blue represents negative value of averaged magnetization, $\left<M_z\right>$, indicating switching of a core, while red region represents no switching. 
\label{Fig:2}}
\end{figure}
%%%%
Next we study the magnetic vortex core switching by numerically solving the Landau-Lifshitz-Gilbert (LLG) equation including $H_{\rm{TIFE}}$  by using the forth-order Runge-Kutta method.
As seen from Eq. (\ref{eq:a-SC}), the effective field can not induce magnetization dynamics in zero temperature because the field direction is perfectly parallel (or anti-parallel) to the magnetization.
The dynamics emerges when we include the thermal fluctuation of the magnetization, similarly to the case of the conventional IFE on a uniform ferromagnet\cite{rf:Kirilyuk,rf:Radu11}.
The thermal fluctuation is included as a random field, $\zeta$, whose ensemble average is defined by a fluctuation-dissipation theorem as $\langle \zeta_i \rangle_\zeta =0 $ and $\langle \zeta_i(t) \zeta_j(t') \rangle_\zeta =  \frac{2 \alpha k_{\rm{B}} T}{\gamma g\mu_{\rm{B}} } \delta_{ij}\delta(t-t')$\cite{rf:Radu11,rf:Garanin97}, where $\gamma$ is the gyromagnetic ratio and $\alpha$ is the Gilbert damping constant.
We also include an internal field, $\bm{H}_{\rm{I}}$, needed to describe a vortex structure, representing the exchange interaction among the local spins and magnetic anisotropy. 
The LLG equation is therefore written as 
\begin{align} \notag
\dot{\bm{n}} = & \gamma \left[ \bm{n} \times \left( \bm{H}_{\rm{TIFE}} + \bm{H}_{\rm{I}} + \bm{\zeta} \right) \right] \\ \label{eq:LLG}
		 & -  \alpha \gamma \left[ \bm{n} \times \left( \bm{n} \times \left( \bm{H}_{\rm{TIFE}} + \bm{H}_{\rm{I}}  \right)  \right) \right].
\end{align}
We consider a permalloy disk with a radius of $R=1$ $\mu$m.
We change the thickness of the disk from $d = 5$ nm to $40$ nm.
A vortex structure in the absence of light is calculated by minimizing the exchange energy and demagnetization energy.
The radius of the vortex core is 25 nm.
A circularly polarized light is irradiated perpendicular to the disk. 
The light-induced magnetic field, $\bm{H}_{\rm{TIFE}}$, is plotted in Fig. 1 (c). Even near a threshold value of $L=0.3 \times 10^{-18} $ Tm$^2$, the maximum filed acting at the center of core reaches 0.3 T, but the filed decays rapidly as distance from the center increases. 
This field at the center is strong enough to induce picosecond core reversal. 
(In Ref. 2, uniform magnetic field of $0.03$ T was used for a flip within 140 ps.)

Figure 2 shows our result of averaged magnetization perpendicular to the vortex plane ({\it i.e.} the polarization of the vortex core) as a function of time and light intensity $L\equiv |\bm{L}|$. A circularly polarized light is irradiated at $t=0$. We see immediately that there is a threshold value for the core flip at $L \simeq 0.3 \times 10^{-18}$ Tm$^2$.
(We defined the switching time $\tau_{\rm sw}$ as a time when the magnetization $\left<M_z\right>$ becomes $-0.1$ in Fig. 2 (see also Fig. 4 ).) The threshold value of $L = 0.3 \times 10^{-18}$ $ $Tm$^{2}$ in Fig. 2 corresponds to the electric field of $E\simeq 6 \times 10^{5}$ V/m, and to the laser intensity of $10 ^{-3}$ J/m$^{2}$ if a pulse duration is 0.25 ns. It is seen from Fig 2 that a quite fast switching time ($\tau_{\rm sw}=150$ ps at $L=0.4 \times 10^{-18}$  $ $Tm$^2$) is realized. Compared with the current-induced vortex core switching, where $\tau_{\rm sw}=20$ ns\ \  \cite{rf:Yamada06}, the present mechanism is 100 times faster. 
Difference of the switching time between current-induced case and TIFE is understood by where there is a  slow precessional motion. For the current-induce core switching process, a slow precession of a core is induced and a switching occurs when the core velocity becomes sufficiently large.
On the other hand, TIFE does not induce such slow precessional motion.

We found that the switching time $\tau_{\rm sw}$  depends much on the thickness $d$ of the disk.
%%%%%%%%%%
\begin{figure}\centering
\includegraphics[scale=0.3]{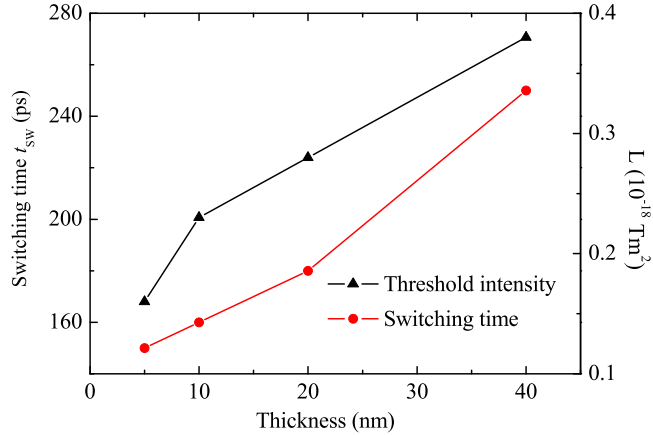}
\caption{ (color online). Thickness dependence of the switching time and the threshold intensity ($L$) for core switching. 
\label{Fig:2}}
\end{figure}
%%%%%%%%%%
As shown in Fig. 3, switching time becomes shorter when the disk becomes thinner. This is because the net magnetization to be reversed is proportional to the thickness. 
In contrast, the switching time is insensitive to the radius of the disk and pulse duration.

From Fig. 2, one also notices that the switching time does not depend much on the applied laser power as long as it is above the threshold.
In fact, even for much high intensity of $L=6\times 10^{-18}$ Tm$^{-2}$, the switching time remains almost the same ($\sim 100 $ ps).
This fact indicates that thermal fluctuation is essential to initiate the switching process, as noticed in a case of uniform ferromagnet\cite{rf:Garanin97}.
In fact, from the inset of Fig. 4, we see that the switching time depends on the temperature logarithmically.
In the present TIFE, therefore, the polarized light intensity merely modifies the energy difference between the up and down states of vortex core, but does not affect the switching speed.   
This is because the magnetic field induced by the incident light, Eq. (\ref{eq:IFE-SC}), is proportional to a local spin at each position. The field does not thus produce a torque which initiates the switching, and the initial torque must be induced by thermal fluctuation.

%%%%%%%%%%
\begin{figure}\centering
\includegraphics[scale=0.27]{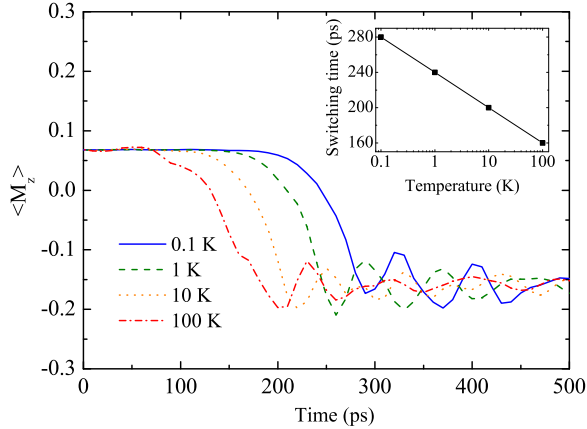}
\caption{(Color online). Time-dependence of the averaged magnetization during the core switching process for several temperatures $T= 0.1$ K, 1 K, 10 K, and 100 K. 
The intensity of the light is $L = 0.6 \times 10^{-18}$ $ $Tm$^{2}$.
Inset: Dependence of switching time to temperature. 
\label{Fig:4}}
\end{figure}
%%%%%%%%%%

In the conventional IFE in uniform ferromagnets,  the switching time is inversely proportional to the temperature\cite{rf:Kirilyuk}. This dependence was explained by deriving an effective LLG equation by taking averaging of thermal field, $\zeta$ in Eq. (\ref{eq:LLG}). The effective Landau-Lifshitz-Gilbert equation for the average magnetic moment, $\bm{m} \equiv \langle \bm{n} \rangle$, reads 
\begin{align} \notag
\dot{\bm{m}} = & \gamma \left[ \bm{m} \times  \left(\bm{H}_{\rm{IFE}} + \bm{H}_{\rm{I}}\right)  \right] 
	 - \tau_{\rm{L}}^{-1} \bm{m} 
	 \\  \label{eq:LLB}
	& - \alpha \gamma \langle \left[ \bm{n} \times \left( \bm{n} \times \left(\bm{H}_{\rm{IFE}} + \bm{H}_{\rm{I}}\right) \right) \right]\rangle_{\zeta}, 
\end{align}
where $\bm{H}_{\rm IFE}$ is the field generated by the conventional IFE and $\tau_{\rm{L}} = \frac{ g \mu_{\rm{B}}}{\gamma \alpha k_{\rm{B}}T}$ is a longitudinal relaxation time\cite{rf:Garanin97}.
Since this relaxation time determines the switching time, the switching time of a uniform magnetization is inversely proportional to the temperature. 	
In contrast to the conventional IFE on a uniform ferromagnet, TIFE for a vortex leads to a weak logarithmic dependence on the temperature. 
This dependence would be understood by noting that spins in a vortex are strongly correlated to each other and thus a collective switching process not expressed by a single relaxation time occurs. 

%Summary
In conclusion, we have discovered an unconventional inverse Faraday effect which appears when a circularly polarized light is irradiated to ferromagnet with a topological spin structure such as vortices and skyrmions.
The effective magnetic field is derived analytically by using the non-equilibrium Green's function.
We have shown by numerical simulation that a ultrafast switching of the magnetic vortex core is possible by use of TIFE resulting in a switching time of 150 ps, 100 times faster than that induced by the electric current.
Even faster switching is expected if heating effect is included. 
Proposed mechanism opens a new way of the read-write process in spintronics devices.

%\section*{Acknowledgments}
K.T thanks Y. Nozaki for useful discussions. This work was supported by a Grant-in-Aid for Scientific Research (B) (Grant No. 22340104) from Japan Society for the Promotion of Science and UK-Japanese Collaboration on Current-Driven Domain Wall Dynamics from JST. K.T is financially supported by the Japan Society for the Promotion of Science for Young Scientists.

\end{document}